\documentclass[t12pt,pra]{revtex4}
\usepackage{graphicx}
\usepackage{epsfig}
\usepackage{latexsym}
\usepackage{amssymb}
\usepackage{bm}
\usepackage{float}
\usepackage{subfigure}
\usepackage{amsmath}
\usepackage{amsfonts}
\begin{document}

\title{From B-Modes to Quantum Gravity and Unification  of Forces\footnote{Essay Written for the 2014 Gravity Research Foundation Awards for Essays on Gravitation}}

\date{March 31, 2014}
\author{Lawrence M. Krauss}
\affiliation{Department of Physics and School of Earth and Space Exploration, Arizona State University, PO BOX 871404 Tempe, AZ 85287-1404, USA \\ and \\ Mount Stromlo Observatory, Research School of Astronomy and Astrophysics, Australian National University, Weston, ACT, Australia, 2611)}
\email{krauss@asu.edu}
\author{Frank Wilczek}
\affiliation{Department of Physics, MIT, Cambridge MA 02139}
\email{wilczek@mit.edu}

\begin{abstract}

It is commonly anticipated that gravity is subject to the standard principles of quantum mechanics.   Yet some -- including Einstein -- have questioned that presumption, whose empirical basis is weak.   Indeed, recently Freeman Dyson has emphasized that no conventional experiment is capable of detecting individual gravitons.   However, as we describe, if inflation occurred, the Universe, by acting as an ideal graviton amplifier, affords such access.   It produces a classical signal, in the form of macroscopic gravitational waves, in response to spontaneous (not induced) emission of gravitons.  Thus recent BICEP2 observations of polarization in the cosmic microwave background will, if confirmed, provide firm empirical evidence for the quantization of gravity.  Their details also support quantitative ideas concerning the unification of strong, electromagnetic, and weak forces, and of all these with gravity.  

\end{abstract}

\maketitle

The possible detection of primordial gravitational waves via B mode polarization in the cosmic microwave background \cite{Ade:2014xna}, if confirmed, will provide a dramatic new window on the early universe, back to a time as early as $10^{-35}$ seconds after the big bang.  Cosmic inflation at a high energy scale is the most plausible source of such primordial waves (e.g. \cite{starobinsky,Rubakov:1982df,Fabbri:1983us,Abbott:1984fp,Krauss:1992ke}, and predicts several aspects of the radiation  that can be checked in future observations.   As a result, the BICEP2 results afford brilliant new opportunities to test the inflationary paradigm \cite{Guth:1980zm,Linde:1981mu}, and to explore other possible dynamics at that time (for a recent discussion see \cite{dentetal}). They will also, of course, represent the first direct observation of gravitational waves.

Here we consider two other major implications of the discovery, concerning fundamental interactions:
\begin{itemize}
\item   Gravitational waves from inflation arise from quantum gravitational fluctuations in the de Sitter-like expansion phase.  Thus their observation directly confirms the treatment of gravity as a quantum theory.  One method of calculating the gravitational waves traces their origin to spontaneous emission of single gravitons, which are then amplified classically, by expansion, into the observed waves.   Thus their observation certainly supports, and arguably proves, the existence of gravitons. 
\item The amplitude of the inflationary gravity wave background is determined by the rate of cosmological acceleration, which in turn is governed by the magnitude of the energy density that drives inflation.  The BICEP2 signal corresponds to an energy scale suggested, on quite independent grounds, by unified field theories of the strong, weak, and electromagnetic interactions.   This scale is significantly, but not vastly, below the Planck scale.  That circumstance implies that the calculation of the unification scale is not complicated by quantum gravity effects, and suggests that gravity might also be brought into the unification.  In detail, quantitative unification of the forces is sensitive to the particle content of the theory, and works best if superpartners to the known particles exist and are not too heavy.  Plausibly, some of those superpartners should be observed as the Large Hadron Collider (LHC) achieves its design energy.   
\end{itemize}

There is considerable synergy among all these conclusions.  Together they suggest that bold but straightforward extrapolation of the known laws of physics, enriched with additional symmetry but not qualitatively modified, to much more extreme conditions of high energies, small distances, and short times than those where they were established, leads to a successful description of Nature.  

\bigskip

It has been appreciated since the early work of Grischuk \cite{grsh} that in linearized gravity the two helicity states associated with propagating gravitational modes each obey the equations of motion for a scalar field in a background expanding space.  If one quantizes these modes in the standard way, with a correspondence between quantized mode functions $h_k$ for gravity, and scalar field mode functions $v_k$ given by

\begin{equation}
h_k={2 \over {aM_{pl}}} v_k
\end{equation}
then it is straightforward to calculate the amplitude of large wavelength zero-point fluctuations in these modes in a background de-Sitter space:
\begin{equation}
\left< h_{\bf k}h_{\bf k'}\right> = {4\over {k^3}} {H^2 \over M_{pl}^2} \delta(\bf{k+k'})
\end{equation}
Once these modes leave the horizon during the inflationary expansion, they freeze in, and expansion effectively amplifies the mode number.  When they later return inside the horizon, they appear as a coherent superposition of many quanta, i.e. as a classical wave.   These waves display a dimensionless power spectrum at the horizon, given by (e.g. \cite{Rubakov:1982df,Fabbri:1983us,Abbott:1984fp,Krauss:1992ke})
\begin{equation}\label{gWaveSpectrum}
\Delta^2(k) = {k^3 \over {2 \pi^2}} P_t = {2\over {\pi^2}} {{H^2 \over M_{pl}^2} }
\end{equation}
In this calculation the initial mode number is small, thus implicating quantum gravity.  
Note that the amplitude of the power spectrum is independent of any inflationary parameters except the expansion rate $H$ during inflation which, by Einstein's equations, depends upon the energy density stored in the inflaton field.   While calculation of scalar field ``inflaton'' fluctuations, which depend on an assumed potential, lead to an eventual prediction of adiabatic scalar energy density fluctuations resulting after inflation, the predicted gravitational wave amplitude is independent of details of the unknown inflationary potential.    
Density perturbations predicted by inflation result in a measurable temperature anisotropy in the Cosmic Microwave Background (CMB) radiation.  Since this anisotropy has been measured, inflationary models are tuned to produce scalar density perturbations in agreement with observation.  

The amplitude of gravity waves is usually reported in terms of $r$, a measure of their ratio to scalar waves. The numerical relation between $r$ and the scale of inflation is 
\begin{equation}\label{scale}
{\cal E}_{\rm inflation} ~=~  1.06 \times 10^{16} {\rm GeV} \left({r \over {0.01}}\right)^{1/4} ~=~ V^{1/4}
\end{equation}
 ({\it e.g}. \cite{baum}) where the final equation reflects that in the simplest inflationary models, involving a single scalar field, the energy scale is the fourth root of the absolute value of the potential (vacuum energy) during inflation.   

For a large enough scale of Inflation, $r$ is sufficiently large to produce observable effects in the CMB.  The simplest effect of an inflationary background of gravitational waves is to produce a direct quadrupole anisotropy in the observed CMB radiation.  It was pointed out as early as 1992 \cite{Krauss:1992ke} that if the inflationary scale was as large as $5 \times 10^{16}$ GeV gravitational waves have could account for the entire observed quadrupole anisotropy.   

A more sensitive probe of the impact of gravitational waves comes from the fact that horizon-sized gravitational waves at the time the CMB was generated will polarize the resulting radiation in a particular manner, distinguishable from polarization induced by density fluctuations \cite{pol,kamion, selj}.  Recently, the BICEP2 collaboration \cite{Ade:2014xna} has reported a non-zero value $ r=0.2^{+0.05}_{-0.07}$ based upon careful measurements of the nature of the polarization of the CMB in a small region of the sky.

While all estimates of inflationary gravitational wave production treat the gravity field quantum-mechanically (see {\it e. g}. \cite{maz}), it seems important to establish that the answer relies on that physics, independent of the method used to calculate it.  (After all, many classical results -- in principle, all of them! -- can be calculated using quantum mechanics.)  An objective measure of this reliance, is the dependence of the result on Planck's constant $\hbar$. 

A simple, and therefore robust, argument from dimensional analysis settles the question \cite{Krauss:2013pha}.

As emphasized above, in the de Sitter limit the inflationary epoch is characterized by a single parameter, the Hubble parameter $H$.  Abstracting $M, L, T$ as dimensions of mass, length, and time, we have 
$\left[ H \right] ~=~ \frac{1}{T}$ (where, by convention, a bracketed quantity represents the dimensional content of that quantity).
A contemporary gravitational wave background that was produced during the inflationary epoch will also involve the gravitational constant $G$.   Since the dimensionless ratio $G \hbar H^2 / c^5$ is small for inflation with a curvature scale less than the Planck length, as suggested by BICEP2 (and required for theoretical control) the lowest-order effect, involving one power of $G$, will dominate.    

To form a dimensionless numerical measure of the strength of the gravitational background, we multiply the energy density { $\rho_{\rm grav.} $ in gravitational radiation after inflation ends by a factor $a^4$, where $a$ is the scale factor of the expanding universe after the end of inflation.  This takes into account the redshift of gravitational waves once they enter inside the horizon.  
Thus we must combine $G$ to the first power, together with powers of $H$ and the fundamental constants $\hbar, c$, and $L^4$, to produce an dimensionless invariant measure of the magnitude of the gravity wave background.  The dimensional analysis runs  
\begin{equation}
\left[ G \right] \left[ H \right]^\alpha \left[ \hbar \right]^\beta \left[ c \right]^\gamma ~=~ [ \rho_{\rm grav.} ] [a^4 ]  ~=~  \frac{\left[ E \right]}{L^3} \, L^4 ~=~ \frac{ML^3}{T^2}
\end{equation}
This has a unique solution $\alpha = 2, \beta = 2, \gamma = -4$.  The result, being proportional to $\hbar^2$, is intrinsically quantum mechanical.   Note that if factors of $\hbar$ and $c$ are made explicit in Eqn.\,(\ref{gWaveSpectrum}), our dimensional analysis is consistent with that microscopic calculation.

\bigskip

As seen from Eqn.\,(\ref{scale}), the BICEP result suggests a scale of inflation of ${\cal E}_{\rm inflation} \approx 2 \times 10^{16}$ GeV.   Remarkably, that same scale appears in quite a different context, namely that of unified field theories \cite{dimop}.  (\cite{drw} provides an accessible conceptual review, with extensive references, and \cite{expt} reviews the comparison with experiment.)   

The standard model gauge theories of strong, weak, and hypercharge interactions have provided an accurate, economical, and richly detailed account of all relevant observations.  They also smoothly accommodate gravity, in the form of minimally coupled general relativity.  (Though the equations break down at energies $E \geq M_{\rm Planck} \approx 2 \times 10^{18}$ GeV, and for the corresponding distances, curvatures, etc.)   Since these gauge theories share a common mathematical structure, and are all based on symmetry, it is natural to enquire whether they might all arise as different aspects of a single, all-embracing gauge theory.  The enhanced gauge symmetry contains the $SU(3) \times SU(2) \times U(1)$ symmetry of the strong, weak, and hypercharge interactions as a subgroup.  A crucial first test of this idea is that the quarks and leptons of each family, which appear within 6 separate multiplets of $SU(3)\times SU(2)\times U(1)$ and display peculiar fractional hypercharges, fit into a smaller number of complete mutiplets under the unified symmetry.  $SO(10)$ turns out to work spectacularly well.  It fits all the quarks and leptons of each family into a single irreducible (16-dimensional spinor) representation, and also accounts for their peculiar hypercharges.  

The mathematics of nonabelian symmetry requires that the fundamental coupling constants of the three subtheories in $SO(10) \rightarrow SU(3) \times SU(2) \times U(1)$ are equal.   As observed, they are not.    Quantum field theory teaches us, however, that the perceived strength of a coupling depends on the characteristic energy, or equivalently (inverse) distance, of the process used to measure it.  The running of the couplings, or renormalization, occurs due to the effect of quantum fluctuations in matter fields (also known as virtual particles).   This feature of quantum field theory has been verified quantitatively in innumerable experiments, with considerable precision.  Thus the correct formulation of the predicted unification is not simply that the observed couplings are equal, but rather that their values become equal when extrapolated to energy scales at or above the scale of the condensate.  For at such large scales, the symmetry breaking effect of the condensate will be overcome.  

The required extrapolation depends, in its numerical details, upon the assumed particle content of the unified world-model.  If we use only the known particles, the three different $SU(3)\times SU(2) \times U(1)$ couplings have a near approach, but fail to unify.  But if we include, in addition, precisely the additional vacuum polarization contributed by hypothetical superpartners of the known particles, in a minimal extension of the standard model to include supersymmetry, then accurate unification results.  It occurs at the energy
${\cal E}_{\rm unified} \approx 2 \times 10^{16} {\rm GeV} \ \approx {\cal E}_{\rm inflation}$!

The required superpartners cannot be too heavy, for else their contribution is suppressed.   Some of them ought to become visible at the Large Hadron Collider, once it achieves the higher energies expected in coming years.   Their observation would consummate a  grand synthesis of ideas. 

\bigskip

A classic challenge in fundamental physics is to understand the grotesque smallness of the observed force of gravity, compared to other interactions, as it operates between fundamental particles .   Famously, the gravitational interaction is $\sim 10^{42}$ times smaller than 
any of the other forces.   Again, however, proper comparison requires that we specify the energy scale at which the comparison is made.   Since the strength of gravity, in general relativity, depends on energy directly, it appears hugely enhanced when observed with high-energy probes.  At the scale of unification ${\cal E}_{\rm unified} \sim 2\times 10^{16} {\rm GeV} \ \sim {\cal E}_{\rm inflation}$ the discrepant factor $10^{42}$ is reduced to $\sim 10^4$.  While this does not meet the challenge fully, evidently it marks a big step in the right direction.  Taken together with our earlier result on quantization, it makes a compelling case that gravity should not be considered as a separate phenomenon, but rather as a force on the same footing as, and intimately related to, the others.

\end{document}